%Paper: hep-ph/9210274
%From: "R. Sekhar Chivukula" <sekhar@weyl.BU.EDU>
%Date: Thu, 29 Oct 92 10:45:18 -0500

%%%%%%%uses harvmac.tex%%%%%%%%%%%%%%%%%%%%%%%%%%%%%%%%
\input harvmac
\noblackbox
\def\lae{\raise-.5ex\vbox{\hbox{$\; <\;$}\vskip-2.9ex\hbox{$\; \sim\;$}}}
\def\gae{\raise-.5ex\vbox{\hbox{$\; >\;$}\vskip-2.9ex\hbox{$\; \sim\;$}}}
\def\slash#1{\raise.15ex\hbox{/}\kern-.57em #1}

\def\two#1{\raise1.35ex\hbox{$\leftrightarrow$}\kern-.88em#1}
\def\lefta#1{\raise1.35ex\hbox{$\leftarrow$}\kern-.61em#1}
\def\righta#1{\raise1.35ex\hbox{$\rightarrow$}\kern-.61em#1}
\def\Dslash{\raise.15ex\hbox{/}\kern-.77em D}
\def\np#1#2#3{Nucl. Phys. {\bf #1} (#2) #3}
\def\pl#1#2#3{Phys. Lett. {\bf #1} (#2) #3}
\def\prl#1#2#3{Phys. Rev. Lett. {\bf #1} (#2) #3}

\def\prd#1#2#3{Phys. Rev. D {\bf #1} (#2) #3}
\def\vsl{\raise.15ex\hbox{/}\kern-.57em v}

\def\etal{{\it et al}}

\def\Boxmark#1#2#3{\global\setbox0=\hbox{\lower#1em \vbox{\hrule height#2em
     \hbox{\vrule width#2em height#3em \kern#3em \vrule width#2em}%
     \hrule height#2em}}%
     \dimen0=#2em \advance\dimen0 by#2em \advance\dimen0 by#3em
     \wd0=\dimen0 \ht0=\dimen0 \dp0=0pt
     \mkern1.5mu \box0 \mkern1.5mu }

\Title{\vbox{\baselineskip12pt\hbox{BUHEP-92-36}
\hbox{UCSD/PTH 92-36}\hbox{hep-ph/9210274}}}
{\vbox{\centerline{A Comment on the Strong Interactions of }
        \vskip2pt\centerline{Color-Neutral Technibaryons}}}

\centerline{R. Sekhar Chivukula$^{a,1}$, Andrew G. Cohen$^{a,2}$,
Michael Luke$^{b,3}$ and Martin J. Savage$^{b,4}$}

\footnote{}{$^a$Boston University, Department of Physics,
590 Commonwealth Avenue, Boston, MA 02215}

\footnote{}{$^b$Department of Physics 0319, University of California,
San Diego, 9500 Gilman Drive, La Jolla, CA 92093-0319}

\footnote{}{$^1$sekhar@weyl.bu.edu} \footnote{}{$^2$cohen@andy.bu.edu}
\footnote{}{$^3$luke@yukawa.ucsd.edu} \footnote{}{$^4$savage@thepub.ucsd.edu}

\vskip .2in
\centerline
{\bf ABSTRACT}

We estimate the cross section for the scattering of a slow, color-neutral
technibaryon made of colored constituents with nuclei. We find a cross section
of order $A^2\ 10^{-45}$ cm$^2$, where $A$ is the atomic number of the
nucleus.  Even if technibaryons constitute the dark matter in the galactic
halo, this is too small to be detected in future underground detectors.

\Date{10/92}

If technicolor \ref\techn{S.~Weinberg, \prd {19} {1979} {1277}\semi
L.~Susskind, \prd {20} {1979} {2619}.} is responsible for electroweak symmetry
breaking, then technibaryons are an attractive ``cold dark matter'' candidate
\ref\shmueli{S. Nussinov, \pl {B165} {1985} {55}.} \ref\walker{R.~S.~Chivukula
and T.~Walker, \np {B329} {1990} {445}.} since their numbers relative to
ordinary baryons may be {\it dynamically} determined \ref\barr{S.~Barr,
R.~S.~Chivukula, and E.~Farhi, \pl {B241} {1990} {387}.}
\ref\david{D.~B.~Kaplan, \prl {68} {1992} {741}.}. Current searches
\ref\dmsearch{D.~O.~Caldwell, \etal, \prl {61} {1988} {551} \semi S.~P.~Ahlen,
\etal, \pl {B195} {1987} {603}.} for dark matter candidates using underground
double-$\beta$ decay detectors are able to exclude such technibaryons (with a
mass of order 1 TeV) if they carry a weak charge and are also present with a
density sufficient to explain the galactic halo. This exclusion is the result
of coherent weak interactions in the scattering of the hypothetical
technibaryon with nuclei.

It is nevertheless possible \shmueli\ that the lightest technibaryon has
weak-isospin zero in which case the coherent weak interactions which produce a
large scattering cross section with ordinary matter will be absent.  For
example, in the one-family model \ref\farhi{E. Farhi and L.  Susskind, \prd
{20} {1979} {3404}.} with $N_{TC} = 4$ the lightest technibaryon may have a
techniquark composition of either $UUDE$ or $UDDN$, where $U \choose D$ and $N
\choose E$ are techniquark and technilepton doublets respectively. In this
case \shmueli, the lightest technibaryon has spin and isospin zero and is both
color and electrically neutral; such a particle has no coherent weak
interactions and would therefore be very hard to detect.

Nussinov \ref\shmuelii{S.~Nussinov, \pl {B279} {1992} {111}.} has noted that
if the neutral technibaryon has colored constituents, then at distances less
than approximately one Fermi (where one may ignore issues of confinement) it
will have a nonzero chromoelectric polarizability.  Using a non-relativistic
quark model, Nussinov estimated the cross section for the scattering of a
neutral technibaryon from the chromoelectric fields in a nucleon to be of
order $A^2\ 10^{-38}$ cm$^2$, where $A$ is the atomic number of the nucleus.
While this cross section is too small to be observed currently, it is hoped
that it may be large enough to be observed in future detectors.

In this note, we re-calculate the cross section for the scattering of a
color-neutral technibaryon from nuclei without recourse to the quark model.
For definiteness, we consider the spin-0 technibaryon $\phi$ of the one-family
model discussed above.

Consider constructing an effective Lagrangian to describe the interaction of
$\phi$ with the color field. The leading interaction comes from the two
dimension seven operators
\eqn\ei{ {{g^2 \over (4\pi f)^3}} \left(c_1\phi_v^* \phi_v G^{a\mu\nu}
G^a_{\mu\nu}+c_2 \phi_v^*\phi_v
G^{a\mu\alpha}G^{a\nu}_\alpha v_\mu v_\nu\right)\ ,}
where $G^a_{\mu\nu}$ is the color field strength tensor, $g$ is the color
coupling constant, and $f$ is the analog of $f_\pi$ for the technicolor theory
(which is approximately 125 GeV in the one-family model).  Since the
technibaryon is much heavier than $\Lambda_{\rm QCD}$, strong interactions do
not change its velocity.  It is then convenient to describe technibaryons with
different velocities by distinct fields \ref\georgi{H.~Georgi,
\pl{240}{1990}{447}.}.  $\phi_v$ is related to the usual field $\phi$ by the
field redefinition
\eqn\redefn{\phi_v(x)=\sqrt{2m_\phi}e^{im_\phi v_\alpha x^\alpha}\phi(x),}
where $v^\mu$ is the four-velocity of the technibaryon.
Because of the phase redefinition \redefn, derivatives acting on $\phi_v$
bring down factors of the small ``residual'' momentum $k^\mu\equiv
p^\mu-m_\phi v^\mu$.  This prevents the derivative expansion from containing
potentially troublesome terms proportional to $m_\phi v^\mu/4\pi f\sim
\CO(1)$.

Classically, the interaction \ei\ corresponds to the interaction of induced
chromoelectric and chromomagnetic dipole moments with the color field.  Here,
$c_1$ and $c_2$ are unknown strong interaction constants which, according to
the rules of dimensional analysis \ref\nda{ H.  Georgi and A. V. Manohar, \np
{B234} {1984} {189}.}, are expected to be of order 1. The operators and the
coupling $g$ are all understood to be renormalized at a scale of order 1 TeV.
The matrix element of the second operator in \ei\ in nuclei may be related to
the structure function conventionally called $F_1$; however it is suppressed
relative to the first operator in \ei\ by a factor of $\sim\alpha_s(1\ {\rm
TeV})$, and we will consequently neglect it.

Up to small corrections of order $\alpha^2_s(1\ {\rm TeV})$, we may rewrite
the first term in \ei\ as
\eqn\eii{-c_1 {\beta(g) \over 4 \pi f^3 b g} \phi_v^* \phi_v G^{a\mu\nu}
G^a_{\mu\nu}\ ,}
where
\eqn\eiii{ \beta(g) \approx {-b g^3 \over 16 \pi^2}}
is the beta function for the color coupling $g$.  This reformulation is useful
since the operator ${\beta \over g} G^2$ is a renormalization group invariant
and consequently its matrix elements between physical states are
renormalization scale independent.

\relax To calculate the cross section for scattering from a nucleon, we must
evaluate the matrix element
\eqn\eiv{ \langle k,s | {\beta \over g} G^2(q) | k^\prime,s \rangle
\equiv -4 m^2 S(q^2) ,}
where $|k,s\rangle$ is a nucleon state (either $p$ or $n$) of momentum $k$ and
spin $s$, $q=k-k^\prime$ is the momentum transfer, $m$ is the nucleon mass and
$S(q^2)$ is a form factor.  As noted by Voloshin and Zakharov
\ref\vz{M.~Voloshin and V.~Zakharov, \prl{45}{1980}{688}.}, up to corrections
suppressed by powers of the light quark masses, the operator ${\beta \over g}
G^2(0)$ is related to the generator of scale transformations; as noted in
\ref\ccgm{R.S.~Chivukula, A.G.~Cohen, H.M.~Georgi and A.V.~Manohar \pl {B222}
{1989} {258}.} this amounts to differentiating with respect to the logarithm
of $\Lambda_{\hbox{QCD}}$. The result is that, at zero momentum transfer
\eqn\eeiv{ \langle k,s | {1\over 2}{\beta \over g} G^2(0) | k^\prime,s \rangle
= - 2m^2\ .}
This implies $S(0)=1$.

For collisions of technibaryons in the galactic halo (which have a velocity
$v_{halo} \approx 10^{-3} c$) with nuclei in underground detectors, the
momentum transfer is of order a few tens of MeV/c, while the scale of momentum
over which $S$ varies is of order 100 MeV/c \ref\nucl{see, for example, J.
Engel, S. Pittel and P. Vogel, Bartol Preprint BA-92-23.}. Therefore, to a
good approximation, we may replace $S(q^2)$ with $1$. If the momentum transfer
increases to a few hundred MeV/c this form factor will further suppress the
cross section for elastic scattering.

We may now calculate the cross section for the scattering of a slow
technibaryon from a nucleus in an underground detector
\eqn\crsec{\sigma \simeq {c_1^2 A^2 m^4 \over (4\pi)^3 (1+Am/M)^2 b^2 f^6 }\ ,}
where $M$ is the mass of the technibaryon and $A$ is the atomic number of the
nucleus. Here we have assumed that the technibaryon scatters coherently from
all of the nucleons in a nucleus. Taking $m=1$ GeV, and $b=7$ (appropriate for
six light quarks) we find a cross section $\sigma \approx A^2\ 10^{-45}$
cm$^2$, independent of the technibaryon mass for $Am/M\ll 1$.  Unfortunately,
this cross section is too small to be observed in any foreseeable future
detector.

Our calculation gives a result approximately seven orders of magnitude smaller
than the estimate given in \shmuelii. This discrepancy can be explained by the
fact that the large quark-model result is due to contributions from regions
where the technibaryon is a distance of order (1 TeV)$^{-1}$ from a quark.
However, this is a distance scale much smaller than the Compton wavelength of
a ``constituent'' quark and the non-relativistic quark model is inappropriate.
By contrast, since we are able to relate the cross section to the matrix
element of ${\beta \over g} G^2$, our calculation should be trustworthy.

Two issues remain. Firstly, we have so far only considered the coupling of our
neutral technibaryon to gluons, without direct couplings to the light quarks.
Such couplings will appear in our effective theory, but they will be
suppressed either by powers of the extended technicolor scale \ref\el{E.
Eichten and K. Lane, \pl {90B} {1980} {125}\semi S. Dimopoulos and L.
Susskind, \np {B155} {1979} {237}.} or by powers of $\alpha_s(1\ {\rm TeV})$.
For an ETC scale bigger than a few times $10$ TeV, these contributions are
negligible compared to the gluonic piece we have calculated.  Secondly, in
other models of technicolor the lightest technibaryon may be a fermion rather
than a boson.  This makes no difference since the non-relativistic scattering
amplitude will not distinguish a fermion from a boson.
\bigskip
\noindent{\bf Acknowledgements}
\medskip
We would like to thank Shmuel Nussinov for useful conversations and comments.
R.S.C. and M.J.S. thank Robert Jaffe and Emil Mottola for organizing the {\it
Sante Fe Workshop on Hadrons and Physics Beyond the Standard Model} where some
of this work was completed.  R.S.C.  acknowledges the support of an Alfred P.
Sloan Foundation Fellowship, an NSF Presidential Young Investigator Award, a
DOE Outstanding Junior Investigator Award, and a Superconducting Super
Collider National Fellowship from the Texas National Research Laboratory
Commission. A.C acknowledges the support of a DOE Outstanding Junior
Investigator Award.  M.J.S. acknowledges the support of a Superconducting
Supercollider National Fellowship from the Texas National Research Laboratory
Commission.

This work was supported in part under NSF contracts
PHY-9057173 and PHY-8958081 and DOE contracts DE-AC02-89ER40509,
DE-FG02-91ER40676 and DE-FG03-90ER40546, and by
funds from the Texas National Research Laboratory Commission under grants
RGFY91B6 and FCF9219.

\listrefs
%\listfigs
\bye